\documentclass[twocolumn,secnumarabic,amssymb,nobibnotes,aps,pra]{revtex4-1}

\usepackage[pdftex]{graphicx}
\DeclareGraphicsExtensions{.pdf,.png}

\usepackage[english]{babel}
\usepackage{blindtext}
\blindmathtrue

\usepackage{url}
\usepackage{siunitx}
\usepackage{hyperref}
\usepackage{amsmath}
\usepackage{enumerate}
\usepackage[normalem]{ulem}
\usepackage{algorithm}
\usepackage{algpseudocode}

\usepackage{fancyvrb,color}

\usepackage{xcolor}

\usepackage{xfrac}

\newcommand*\around{{\raise.17ex\hbox{$\scriptstyle\mathtt{\sim}$}}}
\newcommand{\figref}[1]{Fig.~\ref{fig:#1}}

\newcommand{\equref}[1]{Eq.~(\ref{eq:#1})}

\newcommand{\Tabref}[1]{Table~(\ref{Tab:#1})}

\usepackage[draft,nomargin,inline,author=]{fixme}
\fxsetup{theme=color}
\definecolor{fxwarning}{rgb}{0.8,0.0000,0.0000}
\RequirePackage{footnote}

\usepackage[caption=false, font=footnotesize]{subfig}

\usepackage[inline]{enumitem}

\usepackage{booktabs}

\begin{document}

\title{Digital Electronics and Analog Photonics for Convolutional Neural Networks (DEAP-CNNs)}%

\author{Viraj Bangari}%
\email{viraj.bangari@queensu.ca}
\affiliation{Department of Physics, Engineering Physics \& Astronomy, Queen's University, Kingston, ON KL7 3N6, Canada}

\author{Bicky A. Marquez}
\affiliation{Department of Physics, Engineering Physics \& Astronomy, Queen's University, Kingston, ON KL7 3N6, Canada}

\author{Heidi B. Miller}
\affiliation{Department of Physics, Engineering Physics \& Astronomy, Queen's University, Kingston, ON KL7 3N6, Canada}

\author{Alexander N. Tait}
\affiliation{National Institute of Standards and Technology (NIST), Boulder, Colorado 80305, USA}

\author{Mitchell A. Nahmias}
\affiliation{Department of Electrical Engineering, Princeton University, Princeton, NJ 08544, USA}%

\author{Thomas Ferreira de Lima}
\affiliation{Department of Electrical Engineering, Princeton University, Princeton, NJ 08544, USA}%

\author{Hsuan-Tung Peng}
\affiliation{Department of Electrical Engineering, Princeton University, Princeton, NJ 08544, USA}%

\author{Paul R. Prucnal}
\affiliation{Department of Electrical Engineering, Princeton University, Princeton, NJ 08544, USA}%

\author{Bhavin~J.~Shastri}
\email{shastri@ieee.org}
\affiliation{Department of Physics, Engineering Physics \& Astronomy, Queen's University, Kingston, ON KL7 3N6, Canada}

\date{\today}%

\begin{abstract}
Convolutional Neural Networks (CNNs) are powerful and highly ubiquitous tools for extracting features from large datasets for applications such as computer vision and natural language processing. However, a convolution is a computationally expensive operation in digital electronics. In contrast, neuromorphic photonic systems, which have experienced a recent surge of interest over the last few years, propose higher bandwidth and energy efficiencies for neural network training and inference. Neuromorphic photonics exploits the advantages of optical electronics, including the ease of analog processing, and busing multiple signals on a single waveguide at the speed of light. Here, we propose a Digital Electronic and Analog Photonic (DEAP) CNN hardware architecture that has potential to be 2.8 to 14 times faster while maintaining the same power usage of current state-of-the-art GPUs.
\end{abstract}

\maketitle

\section{Introduction}
\label{sec:introduction}
The success of CNNs for large-scale image recognition has stimulated research in developing faster and more accurate algorithms for their use. However, CNNs are computationally intensive and therefore results in long processing latency. One of the primary bottlenecks is computing the matrix multiplication required for forward propagation. In fact, over 80\% of the total processing time is spent on the convolution \cite{Li:2016}. Therefore, techniques that improve the efficiency of even forward-only propagation are in high demand and researched extensively \cite{Jaderberg:2014,Goodfellow:2016}.

In this work, we present a complete digital electronic and analog photonic (DEAP) architecture capable of performing highly efficient CNNs for image recognition. The competitive MNIST handwriting dataset\cite{MNIST:2010} is used as a benchmark test for our DEAP CNN. At first, we train a standard two-layer CNN offline, after which network parameters are uploaded to the DEAP CNN. Our scope is limited to the forward propagation, but includes power and speed analyses of our proposed architecture.

Due to their speed and energy efficiency, photonic neural networks have been widely investigated from different approaches that can be grouped into three categories: (1) reservoir computing \cite{Duport:2016,Brunner:2013,Vandoorne:2014,Larger:2012}; reconfigurable architectures based on (2) ring-resonators \cite{PrucnalBook,TaitBroadcast:2014,TaitMrrWB:2016,Tait:2018:mrrMod}, and (3) Mach-Zehnder interferometers \cite{Hughes:2018,Shen:2017}. Reservoir computing in the discrete photonic domain successfully implement neural networks for fast information processing, however the predefined random weights of their hidden layers cannot be modified \cite{Larger:2012}.

 An alternative approach uses silicon photonics to design fully programmable neural networks~\cite{deLima:2019}, using a so-called broadcast-and-weight protocol \cite{TaitBroadcast:2014,TaitMrrWB:2016,Tait:2018:mrrMod}. This protocol is capable of implementing reconfigurable, recurrent and feedforward neural network models, using a bank of tunable silicon microring resonators (MRRs) that recreate on-chip synaptic weights. Therefore, such a protocol allows it to emulate physical neurons. Mach-Zehnder interferometers have been also used to model synaptic-like connections of physical neurons~\cite{Shen:2017}.
The advantage of the former approach over the latter is that it has already demonstrated fan-in, inhibition, time-resolved processing, and autaptic cascadability \cite{Tait:2018:mrrMod}.
The DEAP CNN design is therefore compatible with mainstream silicon photonic device platforms. This approach leverages the advances in silicon photonics that have recently progressed to the level of sophistication required for large-scale integration. Furthermore, this proposed architecture allows the implementation of multi-layer networks to implement the deep learning framework.

Inspired by the work of Mehrabian et al. \cite{Mehrabian:2018}, which lays out a potential architecture for photonic CNNs with DRAM, buffers, and microring resonators, our design goes a step further by considering specific input representation, as well as an example of how an algorithm for tasks such as MNIST handwritten digit recognition can be mapped to photonics. Moreover, we consider summation of multi-channel inputs, multi-dimensional kernels, the limitations on weights being between 0 and 1, and the architecture for the depth of kernel or inputs.

This work is divided in five sections: Following this introduction, 
in section (\ref{sec:convphoto}), we describe convolutions as used in the field of signal processing. Then, we introduce silicon photonic devices to perform convolutions in photonics. Section (\ref{sec:photoniconv}) introduces a hardware inspired algorithm to perform such full photonic convolutions. In Section (\ref{sec:photoniccnn}), we utilize our previously described architecture to build a two-layers DEAP CNN for MNIST handwritten digit recognition. Finally, in section (\ref{sec:efficiencycalcs}), we show an energy-speed benchmark test, where we compare the performance of DEAP with the empirical dataset DeepBench \cite{DeepBench}.
Note, we have made the high level simulator and mapping tool for the DEAP architecture publicly available~\cite{DEAP}.


\section{Convolutions and Photonics}
\label{sec:convphoto}
\subsection{Convolutions Background}

A convolution of two discrete domain functions $f$ and $g$ is defined by:
\begin{equation}
(f*g)[t]=\sum_{t=-\infty}^{\infty}f[\tau]g[t-\tau],
\end{equation}
where $(f*g)$ represents a weighted average of the function $f[\tau]$ when it is weighting by $g[-\tau]$ shifted by $t$.  The weighting function $g[-\tau]$ emphasizes different parts of the input function $f[\tau]$ as $t$ changes.

\begin{figure}[ht]
\centering
\includegraphics[width=0.5\textwidth]{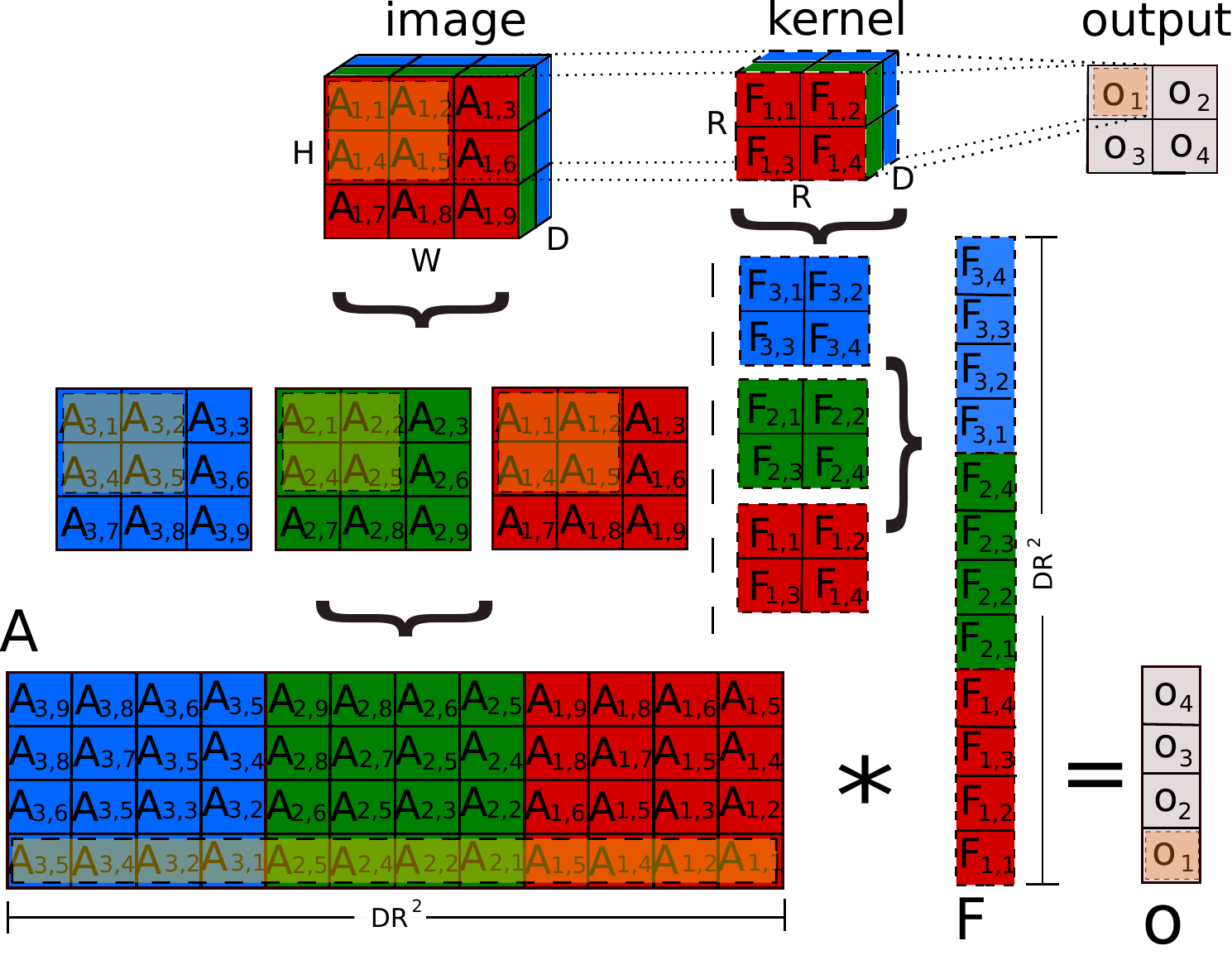}
\caption{Schematic illustration of a convolution. At the top of the figure, an input image is represented as a matrix of numbers with dimensionality $H\times W\times D$ where $H$, $W$ and $D$ are the height, width and depth of the image, respectively. Each element $A_{i,j}$ of $A$ represents the intensity of a pixel at that particular spatial location.
The kernel $F$ is a matrix with dimensionality $R\times R\times D$, where each element $F_{i,j}$ is defined as a real number.
The kernel is slid over the image by using a stride $S$ equal to one. As the image has multiple channels (or depth) $D$, the same kernel is applied to each channel. Assuming $H = W$, the overall output dimensionality is $(H - R + 1)^{2}$.
The bottom of the figure shows how a convolution operation generalized into a single matrix-matrix multiplication.  where the kernel F is transformed into a vector $\mathbf{F}$ with $DR^{2}$ elements, and the image A is transformed into a matrix $\mathbf{A}$ of dimensionality $DR^{2} \times (H - R + 1)^{2}$. Therefore, the output is represented by a vector with $(H - R + 1)$ elements.
} \label{fig1}
\end{figure}

In digital image processing, a similar process is followed. The convolution of an image $A$ with a kernel $F$ produces a convolved image $O$. An image is represented as a matrix of numbers with dimensionality $H\times W$, where $H$ and $W$ are the height and width of the image, respectively. Each element of a matrix represents the intensity of a pixel at that particular spatial location. A kernel is a matrix of real numbers with dimensionality $R\times R$. The value of a particular convolved pixel is defined by:
\begin{equation} \label{eq:val_conv}
O_{i,j}=\sum_{k=1}^{R}\sum_{l=1}^{R}F_{k,l}A_{i+k,j+l}.
\end{equation}
Using matrix slicing notation, \equref{val_conv} can be represented as a dot product of two vectorized matrices:
\begin{equation} \label{eq:conv_gip}
O_{i,j}=\text{vec}(F)^{T}\cdot\text{vec}((A_{m, n})\substack{m\in[i, i+R] \\ n\in[j,j+R]})^{T}.
\end{equation}
A convolution reduces the dimensionality of the input image to $(H-R+1)\times (W-R+1)$, so a padding of zero values is normally applied around the edges of the input image to counteract this.
A schematic illustration of a convolution in digital image processing is shown at the top of Fig.~\ref{fig1}.

\begin{table}[ht]
  \caption{Summary of Convolutional Parameters}
  \begin{center}
    \begin{tabular}{|c|c|}
      \hline
      Parameter & Meaning\tabularnewline
      \hline
      \hline
      $N$  & Number of input images\tabularnewline
      \hline
      $H$ & Height of input image including padding\tabularnewline
      \hline
      $W$ & Width of input image including padding\tabularnewline
      \hline
      $D$ & Number of input channels\tabularnewline
      \hline
      $R$ & Edge length of kernel\tabularnewline
      \hline
      $K$ & Number of kernels\tabularnewline
      \hline
      $S$ & Stride\tabularnewline
      \hline
    \end{tabular}
    \label{Tab:conv_params}
  \end{center}
\end{table}

When convolutions are used to perform parallel matrix multiplications in neural networks such as CNNs, a convolution operation is defined as:
\begin{equation} \label{eq:cnn_conv_def}
O_{i,j}=\text{vec}(F)^{T}\cdot\text{vec}((A_{m, n, k})\substack{m\in[iS, iS+R] \\ n\in[jS, jS+R] \\ k\in[1,D]})^{T},
\end{equation}
where the input $A$ has dimensionality $H\times W \times D$, kernel $F$ has dimensionality $R\times R\times D$ and $D$ refers to the number of channels within the input image. The additional parameter $S$ is referred to as the ``stride'' of the convolution. This convolution is similar to \equref{conv_gip}, except that the outputs from each channel are summed together in the end, and that the stride parameter is always equal to 1 in image processing. The dimensionality of the output feature is:
\begin{equation}
\left\lceil\frac{H-R}{S}+1\right\rceil\times\left\lceil\frac{W-R}{S}+1\right\rceil\times K,
\end{equation}
where $K$ is the number of different kernels applied to an image, and $\lceil \cdot \rceil$ is the ceiling function. \Tabref{conv_params} contains a summary of all the convolutional parameters described so far.

One of the challenges with convolutions is that they are computationally intensive operations, taking up $86\%$ to $94\%$ of execution time for CNNs \cite{Li:2016}. For heavy workloads, convolutions are typically run on graphical processing units (GPUs), as they are able to perform many mathematical operations in parallel.
A GPU is a specialized hardware unit that is capable of performing a single mathematical operation on large amounts of data at once. This parallelization allow GPUs to compute matrix-matrix multiplication at speeds much higher than a CPU \cite{Tan:2011}. The convolution operation can be generalized into a single matrix-matrix multiplication \cite{Chetlur:2014}.
This is shown at the bottom of Fig.~\ref{fig1}, where the kernel $F$ is transformed into a vector $\mathbf{F}$ with dimensionality $KDR^2\times 1$, and the image is transformed into a matrix $\mathbf{A}$ of dimensionality $KDR^2 \times \left\lceil\frac{H-R}{S}+1\right\rceil\left\lceil\frac{W-R}{S}+1\right\rceil K$. Therefore, the output is represented by a vector with $\left\lceil\frac{H-R}{S}+1\right\rceil\left\lceil\frac{W-R}{S}+1\right\rceil K$ elements; where in this particular case $K = 1$, $S = 1$ and $H = W$.

\subsection{Silicon Photonics Background}
An emerging alternative to GPU computing is optical computing using silicon photonics for ultrafast information processing. Silicon photonics is a technology that allows for the implementation of photonic circuits by using the existing complementary-metal-oxide-semiconductor (CMOS) platform for electronics \cite{Rahim:2018}. In recent years, the silicon photonic based “broadcast-and-weight” architecture has been shown to perform multiply-accumulate operations at frequencies up to five times faster than conventional electronics \cite{Nahmias:2018}. Therefore, there is motivation to explore how photonics can be used to perform convolutions, and how it compares to GPU-based implementations.

\begin{figure}[ht]
  \centering
  \includegraphics[width=0.5\textwidth]{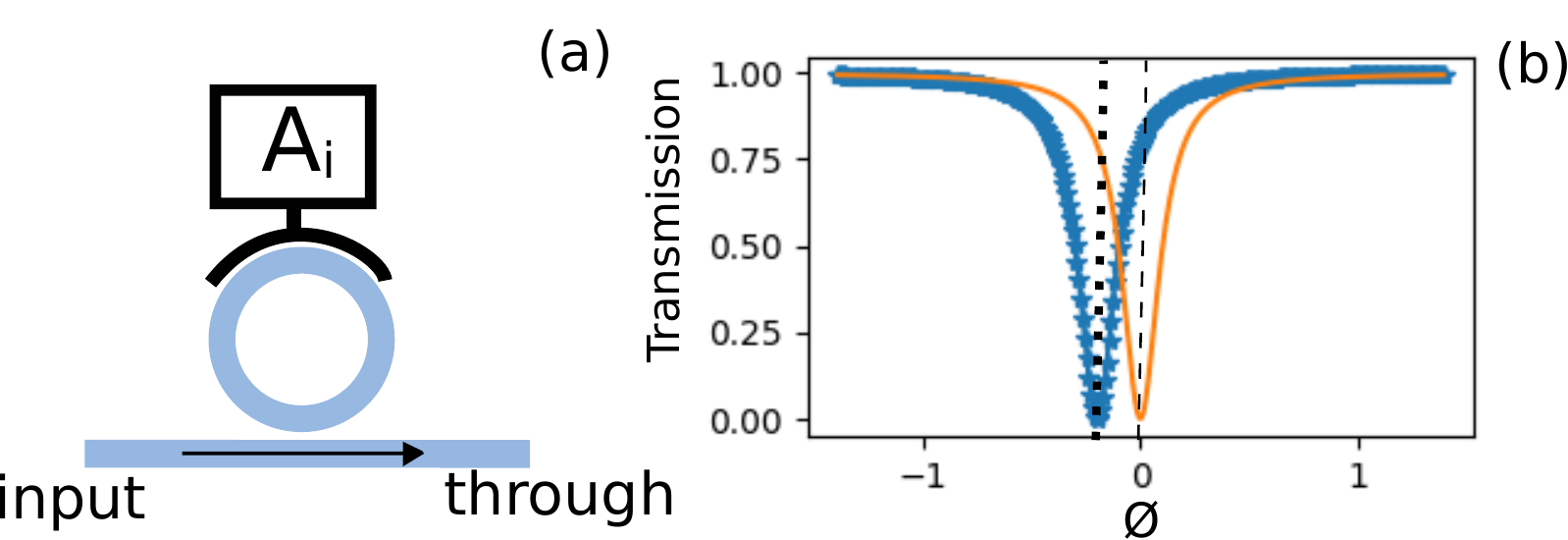}
  \caption{(a) All-pass MRR and (b) transfer function: the orange curve represents the Lorentzian line shape described by Eq.~(\ref{eq1}), centered in the initial phase where MRR is in resonance with the incoming light. The blue triangle curve shows how such phase can be modified by heating the MRR via the application of a current through $A_{i}$.} \label{fig2}
\end{figure}

MRRs are the essential devices of our approach. A MRR is a circular waveguide that is coupled with either one or two waveguides. Such silicon waveguides can be manufactured to have a width  of 500 nm while having a thickness of 220 nm. These waveguides have a bend radius of 5 $\mu$m and can support TE and TM polarized wavelengths between 1.5 $\mu$m and 1.6 $\mu$m \cite{Rahim:2018}. The single waveguide configuration is called an all-pass MRR, see Fig.~\ref{fig2}(a).

The light from the waveguide is transferred into the ring via a directional coupler and then recombined. The effective index of refraction between the waveguide and the MRR and the circumference of the MRR cause the recombined wave to have a phase shift, thereby interfering with the intensity of original light.
The transfer function of the intensity of the light coming out through port with the light going into the input port of the all-pass resonator is described by:
\begin{equation}
T_{n}(\phi) =  \frac{a^2 - 2ra\cos(\phi) + r^2}{1 - 2ra\cos(\phi) + (ar)^2}.
\label{eq1}
\end{equation}
The parameter $r$ is the self-coupling coefficient, and $a$ defines the propagation loss from the ring and the directional coupler. The phase $\phi$ depends on the wavelength $\lambda$ of the light and radius $d$ of the MRR \cite{Bogaerts:2012}:
\begin{equation}
\phi = \frac{4\pi^2 d n_{eff}}{\lambda},
\label{eq2}
\end{equation}
where $n_{eff}$ is the effective index of refraction between the ring and waveguide. The value of $n_{eff}$ can be modified to indirectly change the resonance peak. Such tuning is usually made by applying current to the ring proportional to the variable $A_{i}$. This process heats the ring, yielding a shift of the resonance peak. Figure~\ref{fig2}(b) shows an example of such tuning: the orange curve represents the Lorentzian line shape described by Eq.~(\ref{eq1}), centered in the initial phase of the ring resonator, indicating that the MRR is in resonance with the incoming light. The blue triangle curve shows how such phase can be modified by heating the MRR.

The phase for an all-pass resonator corresponding to a particular intensity modulation value can be computed by using Eq.~(\ref{eq1}):
\begin{equation}
\phi_{i} = \arccos\left[\frac{A_{i}(1+(ar)^2)-a^2-r^2}{2ra(1-A_{i,j})}\right],
\label{eq3}
\end{equation}
resulting in a modulated intensity equal to $A_{i}$:
\begin{equation}
I_{mod} = T_n(\phi_{i})\vert E_0 \vert^2 = A_{i},
\label{eq3_1}
\end{equation}
where $E_0$ is amplitude of the electric field.

An alternative double waveguide configuration is called the add-drop MRR.
The transfer function of the through port light intensity with respect to the input light is:
\begin{equation}
T_{p}(\phi) =  \frac{(ar)^2 - 2r^2\cos(\phi) + r^2}{1 - 2r^2\cos(\phi) + (r^2a)^2};
\label{eq4}
\end{equation}
and the transfer function of the drop port light intensity with respect to the input light is:
\begin{equation}
T_{d}(\phi) =  \frac{(1-r)^2a}{1 - 2r^2\cos(\phi) + (r^2a)^2}.
\label{eq5}
\end{equation}

\begin{figure}[ht]
  \centering
  \includegraphics[width=0.5\textwidth]{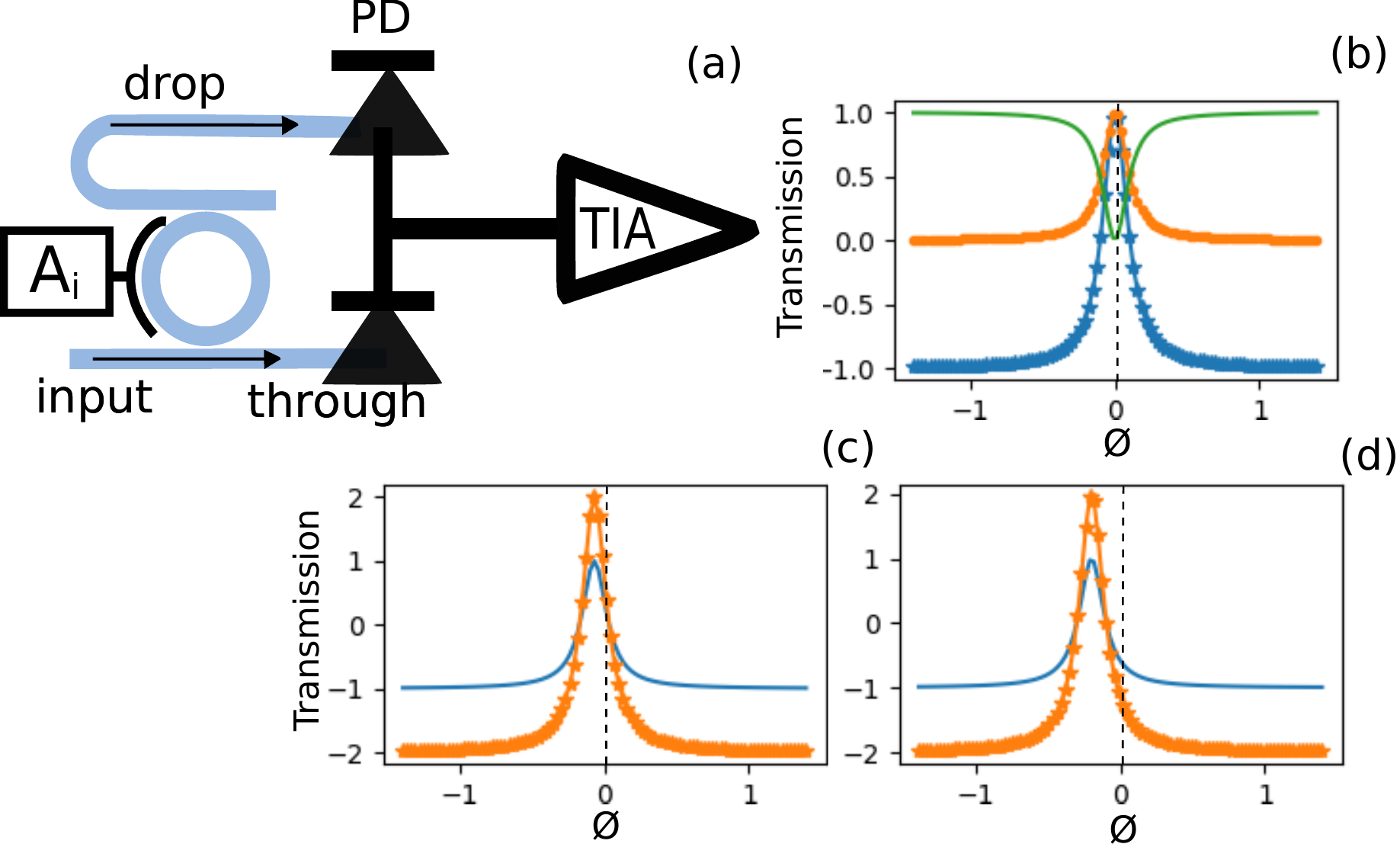}
  \caption{(a) Add-drop configuration and O/E conversion and amplification. (b) Output of the balanced photodiode, the transfer function of $T_{p} - T_{d}$. Orange circle and green curves are drop and through ports, described by Eqs.~(\ref{eq5}) and (\ref{eq4}), respectively. In panels (c) and (d), the phase shifted ($\phi + 0.2)$ blue curves show how such positive and negative kernel values from the drop and the through outputs, respectively. The orange triangle curves show how those values can be amplified by a factor of two using a TIA at the output of the balance photodiode. Those phase shifts are achieved by the application of a current through $A_{i}$. \label{fig3}}
\end{figure}

In the case where the coupling losses are negligible, $a\approx 1$, the relationship between the add-drop through and drop transfer functions is $T_p = T_d -1$. In addition, if we connect the through and drop ports into a balanced photodiode and TIA as in Fig.~\ref{fig3}(a), we get an effective transfer function of $g(T_{p} - T_{d})$ where $g$ is the gain of the TIA. Therefore, we get a modulation of:
\begin{equation}
I_{mod} = g(T_p(\phi_{i}) - T_n(\phi_{i}))\vert E_0 \vert^2 = A_{i}.
\label{eq6_1}
\end{equation}
At the output of the balanced photodiode, the transfer function of $T_{p} - T_{d}$ is shown by the blue triangle curve in Fig.~\ref{fig3}(b). Orange circle and green curves are Lorentzian line shapes, centered in the initial phase where MRR is in resonance with the incoming light, described by Eqs.~(\ref{eq5}) and (\ref{eq4}), respectively.
Differently, Fig.~\ref{fig3}(c) and (d), are centered in a modified phase ($\phi + 0.2$), according to a specific value of the current $A_{i}$.
Here we aim to demonstrate how to represent positive and negative kernel values in analog photonics. This can be achieved by incorporating a balanced-PD at the output of the add-drop MRR. In panels (c) and (d), the blue curves show such positive and negative kernel values from the drop and the through outputs, respectively. The orange triangle curves show the TIA transfer function $g(T_{p} - T_{d})$, where $g$ amplifies $T_{p} - T_{d}$ by a factor of two.

\subsection{Dot Products with Photonics}
The fundamental operation of a convolution is the dot product of two vectorized matrices. Therefore, one needs to understand how to compute a vector dot product using photonics before proposing an architecture capable of performing convolutions.

A wavelength multiplexed signal consists of $k$ electromagnetic waves, each with angular frequency $\omega_{i}$, $i=1,\ldots,k$. If it is assumed that each wave has an amplitude of $E_0$, a power enveloping function $\mu_i$ whose modulation frequency is significantly smaller than $\omega_{i}$, then the slowly varying envelope approximation and a short-time Fourier transform can be used to derive an expression for the multiplexed signal in the frequency domain:

\begin{equation}
E_{mux}(\omega)=\sum_{i=1}^{k}E_{0}\sqrt{\mu_{i}}\delta(\omega-\omega_{i}),
\end{equation}
where $\delta(\omega-\omega_{i})$ is the Dirac delta function and $\mu_{i}\ge0$, since power envelopes are not negative. If the enveloping function is prevented from amplifying the electric field, $\mu_{i}$ can further be restricted to the domain $0\le\mu_{i}\le1$. Next, we introduce tunable linear filters $H^{+}(\omega)$ and $H^{-}(\omega)$ such that when they interact with multiple fields, the following weighted signals are created:

\begin{equation}
\begin{split}
E_{w}^{-}(\omega)&=H^{-}(\omega)E_{mux}(\omega), \\E_{w}^{+}(\omega) &=H^{+}(\omega)E_{mux}(\omega).
\end{split}
\end{equation}

Assuming that the two signals are fed into a balanced photodiode (balanced PD) with spectral response $R(\omega)$, the induced photocurrent is described by:

\begin{equation}
\begin{split}
i_{PD}&=\intop_{-\infty}^{\infty}d\omega R(\omega)\left(\left|E_{w}^{+}(\omega)\right|^{2}-\left|E_{w}^{-}(\omega)\right|^{2}\right), \\&=\intop_{-\infty}^{\infty}d\omega R(\omega)\left(\left|H^{+}(\omega)\right|^{2}-\left|H^{-}(\omega)\right|^{2}\right)\left|E_{mux}(\omega)\right|^{2},\\&=\sum_{i=0}^{k-1}R(\omega_{i})\left(\left|H^{+}(\omega_{i})\right|^{2}-\left|H^{-}(\omega_{i})\right|^{2}\right)E_{0}r_{i}.
\end{split}
\end{equation}

Assuming that $R(\omega)$ is roughly constant in the area of spectral interest, one can set $A_i=E_0R_0\mu_i$ and $F^*_i=\left|H^{+}(\omega_{i})\right|^{2}-\left|H^{-}(\omega_{i})\right|^{2}$ resulting in a photocurrent equal to

\begin{equation}
i_{PD}=\sum_{i=1}^{k}A_iF^{*}_i=\vec{A}\cdot\vec{F}^{*}.
\end{equation}

\begin{figure}[ht]
  \begin{center}
    \includegraphics[width=0.5\textwidth]{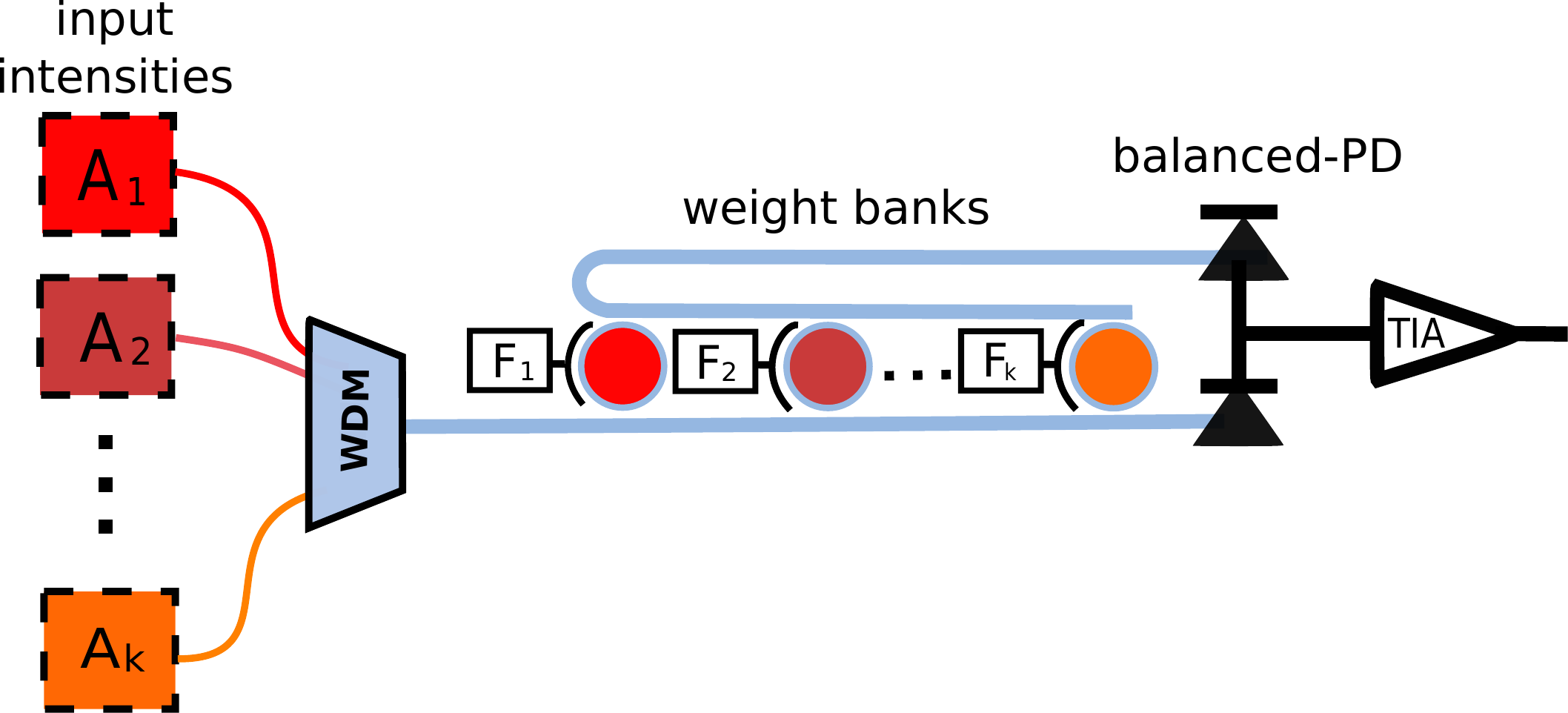}
    \caption[]{An electro-optic architecture that performs dot products. $A_{i}$ ($i=1,\ldots, k$) are input elements encoded in intensities, multiplexed by a WDM and linked to the weight banks via a silicon waveguide. $F_{i}$ are filter values that modulate the MRRs in the PWB. Drop and through output ports are connected to a balanceD-PD, where the matrix multiplication is performed, followed by an amplifier TIA.}
    \label{fig:pwb}
  \end{center}
\end{figure}

\begin{figure*}[ht]
  \centering
  \includegraphics[width=0.9\textwidth]{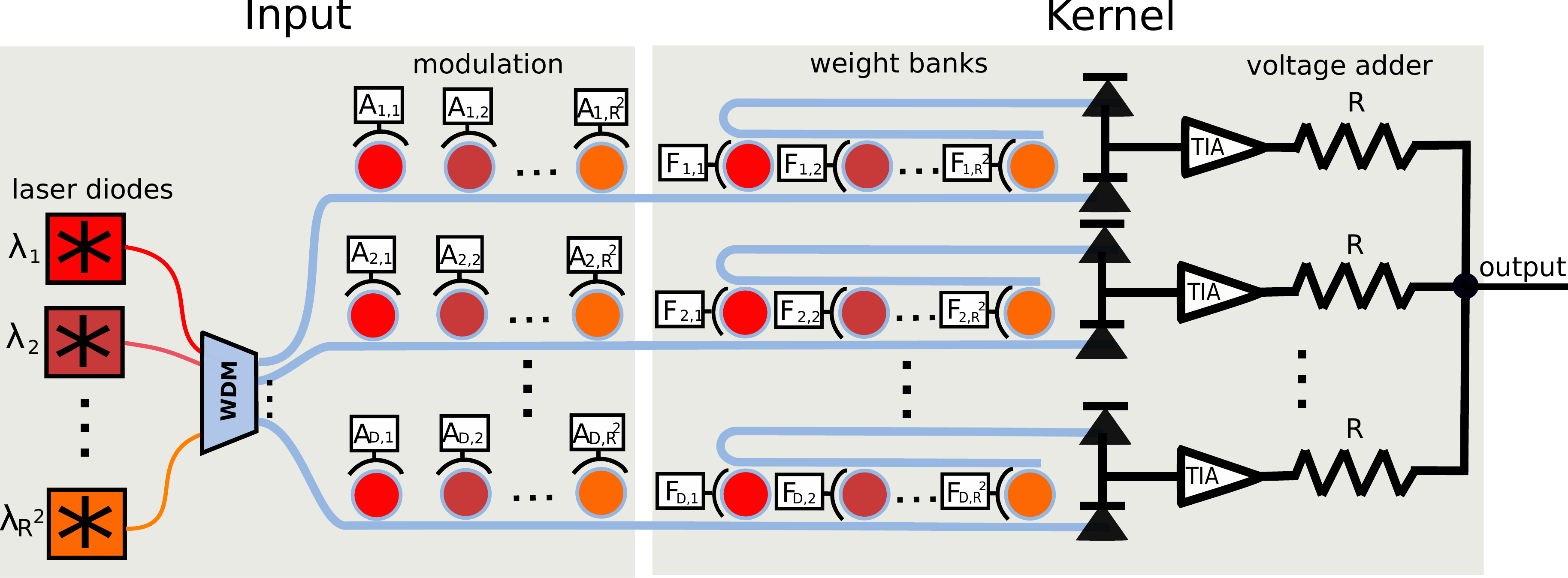}
  \caption{Photonic architecture for producing a single convolved pixel. Input images are encoded in intensities $A_{l,k}$, where the pixel inputs $A_{m,n,k}$ with $m\in[i,i+R_{m}],n\in[j,j+R_{m}],k\in[1, D_m]$ are represented as $A_{l,h}$, $l = 1,\ldots, D$ and $h = 1,\ldots, R^{2}$. Considering the boundary parameters, we set $D = D_{m}$ and $R = R_{m}$. Likewise, the filter values $F_{m,n,k}$ are represented as are represented as $F_{l,h}$ under the same conditions. We use an array of $R^{2}$ lasers with different wavelengths $\lambda_{h}$ to feed the MRRs. The input and kernel values, $A_{l,h}$ and $F_{l,h}$ modulate the MRRs via electrical currents proportional to those values. Once the matrix parallel multiplications are performed, the voltage adder has the function to add all signals from weight banks. Here, $\mathbf{R}$ are resistance values. Then the output is the convolved feature. \label{fig4}}
\end{figure*}

The through and drop ports of a MRR can be used to implement the linear filters $H^+$ and $H^-$ such that $|H^+|^2=T_d$ and $|H^-|^2=T_d$. Knowing that $T_p = T_d - 1$ with minimal losses, we can set a particular weight using:
\begin{equation} \label{eq:weights_i}
F^{*}_i=2T_{d}(\phi_{i})-1,
\end{equation}
where the phase, $\phi_i$ can obtained by using Eq.~\ref{eq4} and Eq.~\ref{eq5} to get:
\begin{equation}
\phi_{i}=\arccos\left[-\frac{1}{2r^{2}a}\left(\frac{2(1-r)^{2}a}{F^{*}_i+1}-1-(r^{2}a)^{2}\right)\right],
\end{equation}
we can see that $F_i^*$ can be between -1 and 1. Since $T_d$ is a filter that only represents values between 0 and 1. In order to perform a dot product with a weight vector $\vec{w}$ whose components are not limited to the range -1 to 1, a gain $g_{TIA}$ can be applied to the photocurrent such that:
\begin{equation}
\begin{split}
\vec{A}\cdot\vec{F}&=g_{TIA}\vec{A}\cdot\vec{F}^{*}\\&=g_{TIA}\sum_{i=1}^{k}A_iF^{*}_i,
\end{split}
\end{equation}
if:
\begin{equation}
g_{TIA}=\max_{1\le i\le k}\left|F_i\right|,
\end{equation}
then,
\begin{equation}\label{eq:gain_tia_weights}
\vec{F}=g_{TIA}\vec{F}^{*};
\end{equation}
assuming that each $\phi_i$ corresponds to a weighting of $w_i^*$. This electronic gain can be performed using a transimpedance amplifier (TIA), which can be manufactured in a standard CMOS process~\cite{Zheng:2017} and packaged or integrated with the photonic chip~\cite{Rahim:2018}. A diagram of the electro-optic architecture described in this section is presented in \figref{pwb}. From now on, this amalgamation of electronic and optical components is referred as a photonic weight bank (PWB). PWBs similar to the one in \figref{pwb} have been successfully implemented in the past \cite{TaitMrrWB:2016,Lipson:2005,TaitFB:2018}.

We can represent negative inputs between -1 and 1 by modifying the power enveloping function to $\mu_i=\frac{1}{2}(x_i+1)$. If the same set of derivations is followed, we can modify \equref{gain_tia_weights} to be:

\begin{equation}
\vec{x}\cdot\vec{w}=g\left(\sum_{i=1}^{k}A_iF^{*}_i+\sum_{i=1}^{k}E_{0}R_{0}F^{*}_i\right).
\end{equation}

The second term in this sum is a predictable bias current term that conceptually be subtracted before feeding into the TIA. This is a disadvantage of supporting negative inputs, as additional optical or electronic control circuitry would need to be designed. Another trade-off is a loss in precision due to a larger range of inputs needing to be represented, analogous to the loss in precision with signed integers for classical computing.

\section{Performing Convolutions using Photonics}
\label{sec:photoniconv}

The goal of this section is to present a photonic architecture capable of performing convolutions for CNNs. This new architecture is called DEAP.

For a maximum number of input channels $D_{m}$ and a maximum kernel edge length $R_{m}$ as bounding parameters for DEAP, we represent the range of convolutional parameters that a particular implementation of DEAP can support. If a convolutional parameter described in \Tabref{conv_params} does not have a complementary bounding parameter, it means that the DEAP architecture can support for arbitrary values of said convolutional parameter.

\begin{figure*}[ht]
  \begin{center}
    \includegraphics[width=0.9\textwidth]{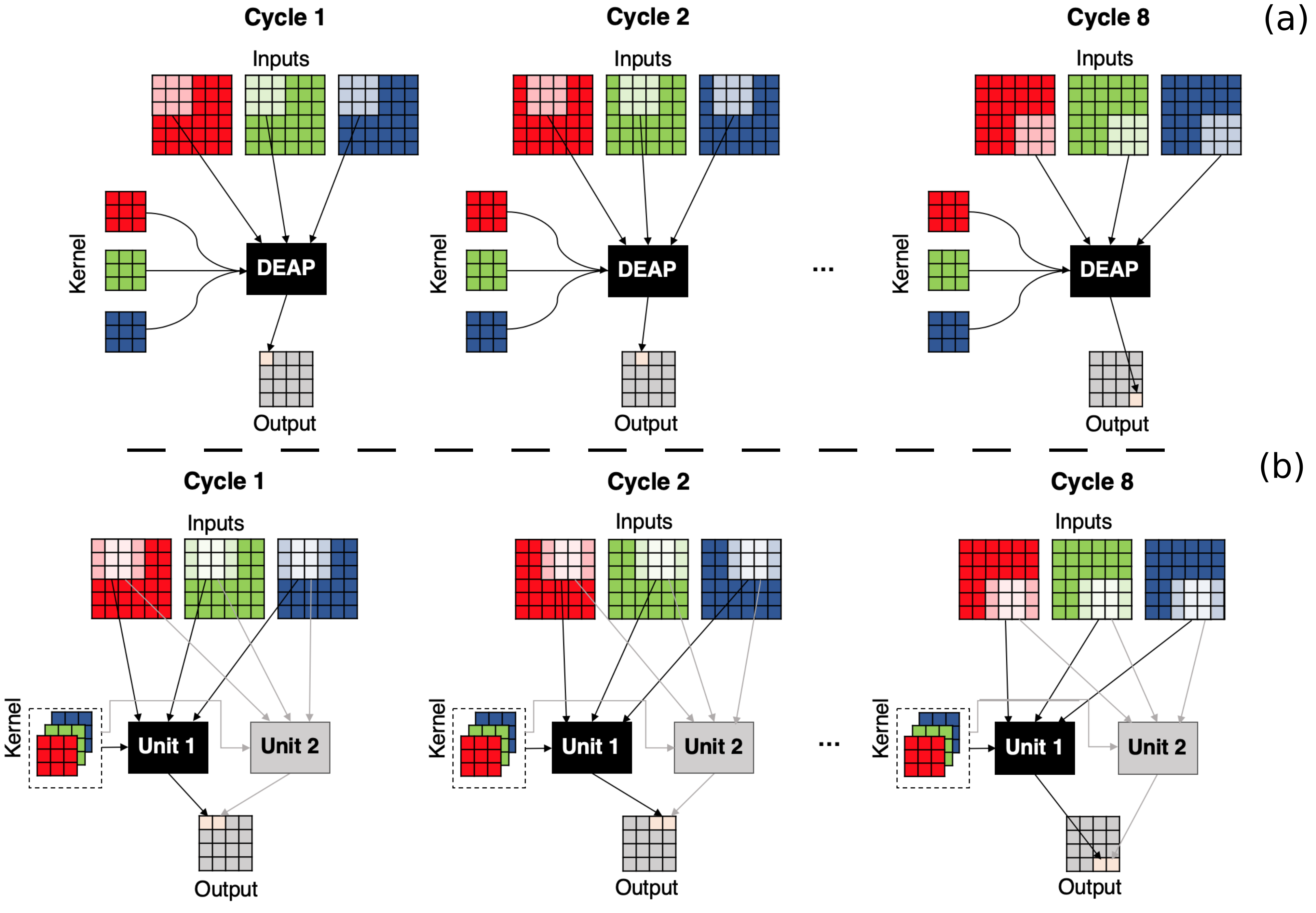}
    \caption[]{(a) Cycling through a convolution using DEAP. (b) Performing a convolution with two convolutional units.}
    \label{fig:cycles}
  \end{center}
\end{figure*}

\begin{figure*}[t]
  \centering
  \includegraphics[width=0.75\textwidth]{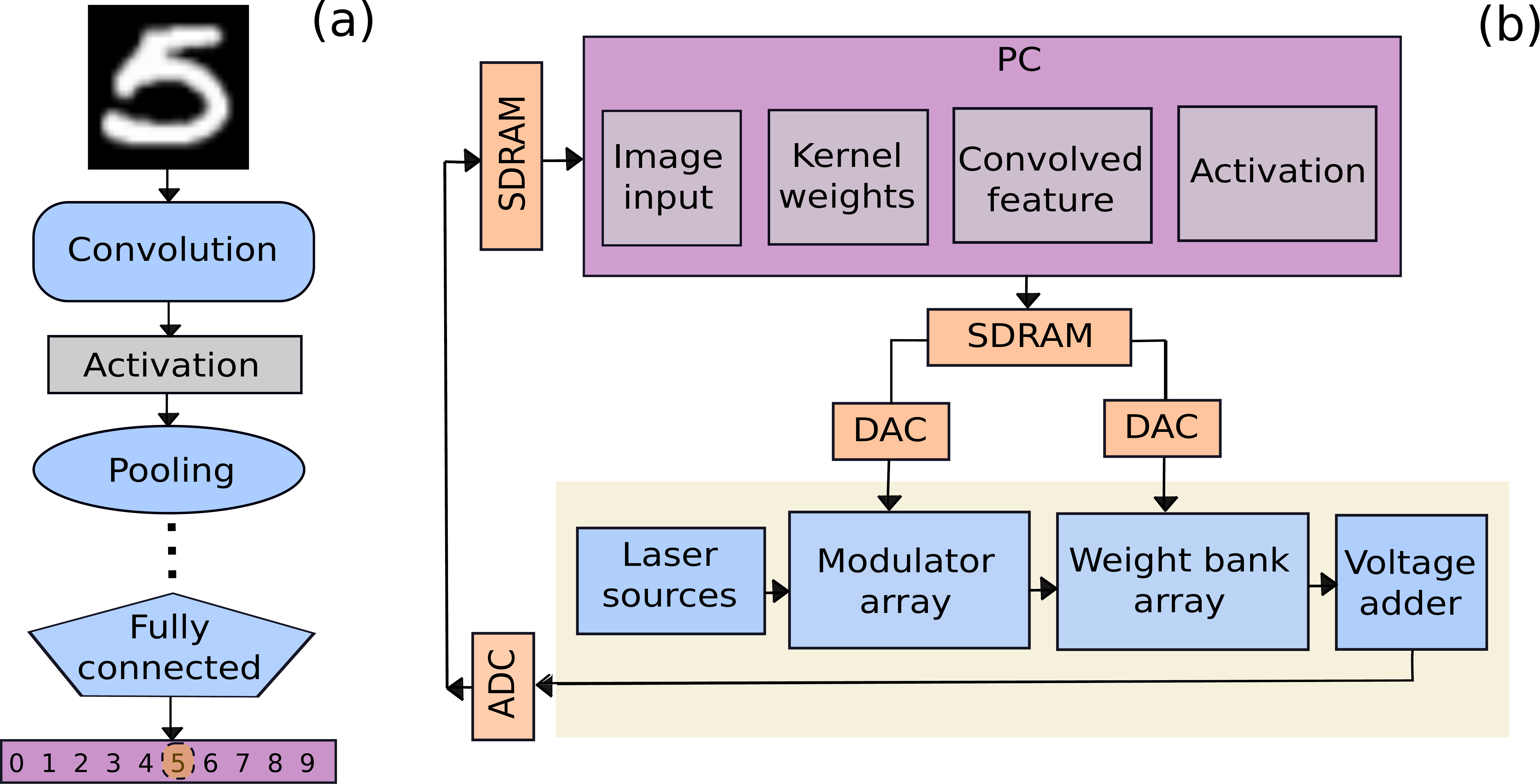}
  \caption{Block diagrams that describe: (a) a typical CNN, which contains convolutions, activation functions, pooling and fully connected layers. In this case we exemplify such diagram using MNIST-based recognition task that predicts the number 5; and (b) the DEAP architecture. In the computer (PC) the input image, kernel weights, and convolved features are stored. Also, the commands to implement the activation function off-chip are stored in the PC. The input image, kernel weights and convolved features are transferred to the chip via DACs from the SDRAM. Then, the convolution is performed on-chip. Finally the output is digitalized via an ADC and stored in a SDRAM connected to the computer.}
    \label{figCNN}
\end{figure*}

\subsection{Producing a Single Convolved Pixel}
First, we consider an architecture that can produce one convolved pixel at a time.
To handle convolutions for kernels with dimensionality up to $R_m \times R_m \times D_m$, we will require $R_m^2$ lasers with unique wavelengths since a particular convolved pixel can be represented as the dot product of two $1\times R_m^2$vectors. To represent the values of each pixel, we require $D_mR_m^2$ modulators (one per kernel value) where each modulator keeps the intensity of the corresponding carrier wave proportional to the normalized input pixel value. The $R_m^2$ lasers are multiplexed together using wavelength division multiplexing (WDM), which is then split into $D_m$ separate lines. On every line, there are $R_m^2$ all-pass MRRs, resulting in $D_mR_m^2$ MRRs in total. Each WDM line will modulate the signals corresponding to a subset of $R_m^2$ pixels on channel $k$, meaning that the modulated wavelengths on a particular line correspond to the pixel inputs $(A_{m, n, k})\substack{m\in[i, i+R_m] \\ n\in[j,j+R_m]}$ where $k\in [1,D_m]$.

The $D_m$ WDM lines will then be fed into an array of $D_m$ PWBs. Each PWB will contain $R_m$ MRRs with the weights corresponding to the kernel values at a particular channel. For example, the PWB on line $k$ should contain the vectorized weights for the kernel $(F_{m, n, k})\substack{m\in[1,R_m^2]\\n\in[1,R_m^2]}$. Each MRR within a PWB should be tuned unique the resonant wavelength within the multiplexed signal. The outputs of the weight bank array are electrical signals, each proportional to the dot product $(F_{m,n,k})\substack{m[1,R_m^2]\\n\in[1,R_m^2]}\cdot (A_{p, q, k})\substack{p\in[i,i+R_m^2]\\q\in[j,j+R_m^2]}$. Finally, the signals from the weight banks need to be added together. This can be achieved using a passive voltage adder. The output from this adder will therefore be the value of a single convolved pixel. Fig.~\ref{fig4} shows a complete picture of what such an architecture would look like.

To perform a convolution with a kernel edge length less than $R_m$, one can set $(F_{m, n, k})\substack{m\in[R+1,R_m]\\n\in[R+1, R_m]}$ to zero. Similarly, if the dimensionality of the kernel is less than $D_m$, then the modulators $(A_{m,n,k})\substack{m\in[1, H]\\n\in[1, W]}$ should also be set to zero, with $k\in[D+1, D_m]$ in this case.

\subsection{Performing a Full Convolution}
In the previous section, we have discussed how DEAP can produce a single convolved pixel. In order to perform a convolution of arbitrary size, one would need to stride along the input image and readjust the modulation array. Since the same kernel is applied across the set of inputs, the weight banks do not need to be modified until a new kernel is applied. \figref{cycles}(a) demonstrates this process on an input with $S=1$. To handle $S\ge 1$, the inputs being passed in to DEAP should also be strode accordingly. In this approach, the inputs should have been zero padded before being passed into DEAP. In pseudocode, performing a convolution with $K$ filters can be implemented as shown in Algorithm \ref{algo_conv_Deap}.

\begin{algorithm}[H]
  \caption{Convolutions for CNNs using DEAP}
  \label{algo_conv_Deap}
  \begin{algorithmic}[1]
  \State $A$ is the input image
  \State $F$ is the kernel
  \State $R$ is the edge length of the kernel
  \State $O$ is a memory block to store the convolution
  \State $S$ is the stride
  \State $H$ and W are the height and width of the input image
  \Function{convolve}{$A, F, R, O, S, H, W$}
    \For{$(k=1;k\le K; k=k+1$)}{}
      \State load kernel weights from F[:,:,:,k]
      \For{$(h=1;h\le H-R+1;h=h+S)$}{}
        \For{$(w=1;w\le W-R+1;w=w+S)$}{}
          \State load inputs from A[h:min(h+T,H), w:min(w+R,W),:]
          \State perform convolution
          \State store results in O[h/S,w/S,k]
        \EndFor
      \EndFor
    \EndFor
  \EndFunction
  \end{algorithmic}
\end{algorithm}

The DEAP architecture also allows for parallelization by treating the photonic architecture proposed in the previous section as a single output ``convolutional unit''. However, by creating $n_{conv}$ instances of these convolutional units, you could produce $n_{conv}$ pixels per cycle by passing in the next set of inputs per unit. This is demonstrated in \figref{cycles}(b) for $n_{conv}=2$. The computation of output pixels can be distributed across each convolutional unit, resulting in a runtime complexity of $O\left(\frac{KHW}{S^{2}n_{conv}}\right)$.

\section{Photonic convolutional neural networks} \label{sec:photoniccnn}

In this section, we show how DEAP can be used to run a CNN. CNNs are a type of neural network that were developed for image recognition tasks. A CNN consists of some combination of convolutional, nonlinear, pooling and fully connected layers \cite{Keiron:2015}, see Fig.~\ref{figCNN}(a). As introduced previously, convolutions perform a highly efficient and parallel matrix multiplication using kernels \cite{Goodfellow:2016}.
Furthermore, since kernels are typically smaller than the input images, the feature extraction operation allows efficient edge detection, therefore reducing the amount of memory required to store those features.

CNNs are networks suitable to be implemented in photonic hardware since they demand fewer resources to do matrix multiplication and memory usage.
The linear operation performed by convolutions allows single feature extraction per kernel. Hence, many kernels are required to extract as many features as possible. For this reason, kernels are usually applied in blocks, allowing the network to extract many different features all at once and in parallel.

In feed-forward networks, it is typical to use a rectified linear unit (ReLU) activation function. Since ReLUs are linear piecewise functions that model an overall nonlinearity, they allow CNNs to be easily optimized during training.
The pooling layer introduces an stage where a set of neighbor pixels are encompassed in a single operation. Typically, such operation consists in the application of a function that determines the maximum value among neighboring values. An average operation can be implemented likewise. Both approaches describe max and average pools, respectively.
This statistical operation allows for a direct down-sampling of the image, since the dimensions of the object are reduced by a factor of two. From this step, we aim to make our network invariant and robust to small translations of the detected features.

The triplet, convolution-activation-pooling, is usually repeated several times for different kernels, keeping invariant the pooling and activation functions. Once all possible features are detected, the addition of a fully connected layer is required for the classification stage. This layer prepares and shows the solutions of the task.

CNNs are trained by changing the values of the kernels, analogous to how feed-forward neural networks are trained by changing the weighted connections \cite{Mehrotra:1996}.  The estimated kernel and weight values are required in the testing stage.
In this work, this stage is performed by our on-chip DEAP CNN. Figure~\ref{figCNN}(b) shows a high-level overview of the proposed testing on-chip architecture. Here, the testing input values stored in the PC modulate the intensities of a group of lasers with identical powers but unique wavelengths. These modulated inputs would be sent into an array of photonic weight banks, which would then perform the convolution for each channel. The kernels obtained in the training step are used to modulate these weight banks. Finally, the outputs of the weight banks would be summed using a voltage adder, which produces the convolved feature. This simulator works using the transfer function of the MRRs, through port and drop port summing equations at the balanced PDs, and the TIA gain term to simulate a convolution. The simulator assumes that the MRRs can only be controlled with $7-$bits of precision as that has been empirically observed in a lab setting.
The MRR self-coupling coefficient is equal to the loss, $r=a=0.99$\cite{Tan:2018} in Eqs.~(\ref{eq1}) (\ref{eq4}) and (\ref{eq5}).

\begin{figure*}[ht]
  \centering
  \includegraphics[width=0.75\textwidth]{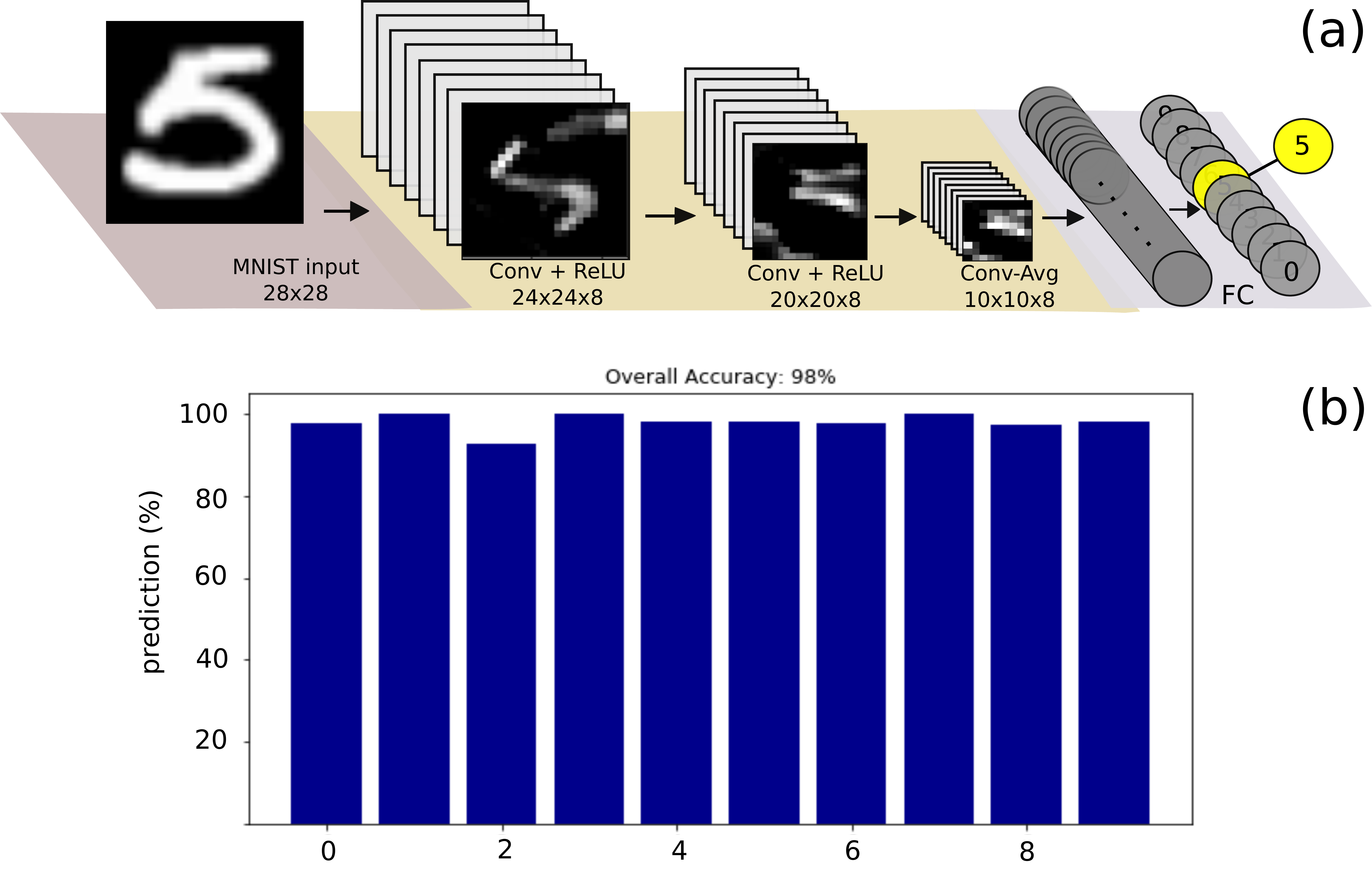}
  \caption{(a) An illustrative block diagram of the two-layers DEAP CNN solving MNIST. (b) Results of the MNIST task using a simulated DEAP CNN.} \label{figRES}
\end{figure*}

The interfacing of optical components with electronics would be facilitated by the use of digital-to-analog converters (DACs) and analog-to-digital converters (ADCs), while the storage of output and retrieving of inputs would be achieved by using memories GDDR SDRAM. The SDRAM is connected to a computer, where the information is already in a digital representation. Then, the implementation of the ReLU nonlinearity and the reuse of the convolved feature to perform the next convolution can be performed.
The idea is to use the same architecture to implement the triplet convolution-activation-pooling on hardware.

In this work, we trained the CNN to perform image recognition on the MNIST dataset. The training stage uses the ADAM optimizer and back-propagation algorithm to compute the gradient function. The optimized parameters to solve MNIST can be categorized in two groups: (i) two $5\times 5\times 8$ different kernels and (ii) two fully connected layers of dimensions $128\times 800$ and $128\times 10$; and their respective bias terms.
These kernels are then defined by eight $5\times 5$ different filters. In the following we use our DEAP CNN simulator to recognize new input images, obtained from a set of 500 images, which are intended to be used for the test step. Our simulator  only works at the transfer level and does not simulate noise or distortion from analog components.
The process of feature extraction performed by the DEAP CNN is illustrated in Fig.~\ref{figRES}(a). As it can be seen in the illustration, a $28\times 28$ input image from the test dataset is filtered by a first $5\times 5\times 8$ kernel, using stride one. The output of this process is a $24\times 24\times 8$ convolved feature, with a ReLU activation function already applied. Following the same process, the second group of filters is applied to the convolved feature to generate the second output, i.e. a $20\times 20\times 8$ convolved feature.

After the second ReLU is applied to the output, average pooling is utilized for invariance and down-sampling of the convolved features.
The average pooling is implemented by a $2\times 2$ kernel whose elements are all $1/4$. However, the stride one was kept; therefore the pooled feature has dimensionality $19\times 19\times 8$. The down-sampling is implemented offline: from the $19\times 19\times 8$ output, a simple algorithm extracts the elements that have even indexes. The result of this process is a $10\times 10\times 8$ pooled output. Finally, the first fully connected layer is fed through by the flattened version of the pooled object. The resultant vector feeds the last fully connected layer, where the result of the MNIST classification appears.

The results of the MNIST task solved by our simulated DEAP CNN is shown by Fig.~\ref{figRES}(b). For a test set of 500 images, we obtained an overall accuracy of $98\%$. This performance was compared to the results obtained using a standard two-layers CNN including a max pooling layer. We found that This standard network achieves an overall accuracy of $98.6\%$. Therefore, we can conclude that our simulator is sufficiently robust despite the $7-$bits of precision considered in the DEAP CNN simulation.

\section{Energy and speed Analyses} \label{sec:efficiencycalcs}

\subsection{Energy Estimation}

The energy used by a single DEAP convolutional unit depends on the $R$ and $D$ parameters. The 100-wavelength limitation for MRRs constrains the maximum $R$ to be 10, as each multiplexed waveguide will store $R^2$ signals. The number of MRRs used in the modulator array is equal to $R^2 D$, meaning that only certain $D$ and $R^2$ values are allowed  for a finite number of MRRs. Assuming that a maximum of 1024 MRRs can be manufactured in the modulator array, a convolutional unit can support a large kernel size with a limited number of channels, $R$ = 10, $D$ = 12, or a small kernel size with a large number of channels, $R$ = 3, $D$ = 113. We will consider both edge cases to get a range of energy consumption values. For the smaller convolution size, we will have $R^2$ lasers, $R^2$ MRRs and DACs in the modulator array, $R^2D$ MRRs and $D$ TIAs in the weight bank array and one ADC to convert back into digital signal. With 100 mW per laser, 19.5 mW per MRR, 26 mW per DAC, 17 mW per TIA \cite{Huang:2016} and 76 mW per ADC, we get an energy usage of 112 W for the large kernel size and 95W for the smaller kernel size. Therefore, we estimate a single convolution unit to use around 100 W when 1024 modulators are used to represent inputs.

\subsection{DEAP Performance}

The time it takes for light to propagate from the WDM to before the balanced PDs is estimated by the following equation:

\begin{equation}
  t_{prop}=\frac{k2\pi r_{MRR}}{c}
\end{equation}

where $c$ is the speed of light $2\pi r_{MRR}$ is the circumference of the MRR and $k$ is the number of MRRs. Assuming 100 MRRs with a radius of 10 µm \cite{TaitMrrWB:2016,Sun:2019}, the PWB gets a propagation time of around 21 ps and a throughput of $1/t_{prop}=50$ GS/s. The bottlenecks come from the fact that the balanced PDs has a throughput of 25 GS/s\cite{Huang:2016} and the TIA has a throughput of 10 GS/s\cite{Atef:2013}. An individual MRR can be modulated at speeds of 128 GS/s\cite{Sun:2019}, meaning that the modulation frequency of the MRRs does not bottleneck the throughput of the PWB.

The throughput of a PWB is around 5 GS/s. The DACs\cite{Sedighi:2012} and ADCs\cite{Fang:2017} both operate at 5 GS/s and support to 7-bits. The GDDR6 SDRAM operates at 16 G with a 256-bit bus size\cite{GDDR6}. Consequently, the speed of the system is limited by the throughput of the DACs/ADCs, resulting in DEAP producing a single convolved pixel at 5 GS/s or $t=200$ ps.

DeepBench \cite{DeepBench} is an empirical dataset that contains how long various types of GPUs took to perform a convolution for a given set of convolutional parameters. Table (\ref{Tab:benchGPUparam}) contains the parameters used for each of these benchmarks, and Table (\ref{Tab:benchGPU}) contains the power consumption.

\begin{table}[ht]
  \caption{Benchmarking parameters for DEAP}
  \begin{center}
  \begin{tabular}{|c|c|c|c|c|c|c|c|}
  \hline
  $W$ & $H$ & $D$ & $N$ & $K$ & $R_{w}$ & $R_{h}$ & $S$\tabularnewline
  \hline
  \hline
  700 & 161 & 1 & 4 & 32 & 5 & 20  & 2\tabularnewline
  \hline
  112 & 112 & 64 & 8 & 128 & 3 & 3 & 1\tabularnewline
  \hline
  7 & 7 & 832 & 16 & 256 & 1 & 1  & 1\tabularnewline
  \hline
  \end{tabular}
  \label{Tab:benchGPUparam}
  \end{center}
\end{table}

\begin{table}[ht]
  \caption{Benchmarked GPUs with power consumption}
  \begin{center}
  \begin{tabular}{|c|c|}
  \hline
  GPU & Power Usage (W)\tabularnewline
  \hline
  \hline
  AMD Vega FE\cite{AMDVEGA} & 375\tabularnewline
  \hline
  AMD MI25\cite{AMDMI25} & 300\tabularnewline
  \hline
  NVIDIA Tesla P100\cite{NVIDIATELSA} & 250\tabularnewline
  \hline
  NVIDIA GTX 1080 Ti\cite{NVIDIAGTX} & 250\tabularnewline
  \hline
  \end{tabular}
  \label{Tab:benchGPU}
  \end{center}
\end{table}

The speeds of various GPUs were directly taken from Ref. \cite{DeepBench}, while the speed of the convolution was estimated using the following equation:

\begin{equation} \label{eq:runtime}
  t_{runtime}=200\text{ ps}\times\frac{NK}{n_{conv}}\left(\frac{H-R}{S}+1\right)\left(\frac{W-R}{S}+1\right).
\end{equation}
In some of the benchmarks, the kernels edge lengths were not equal, hence the parameters $R_w$ and $R_h$ which correspond to the width and height of the kernels. For each of the selected benchmarks, the parameters $R^2 D \le 1024$, meaning that the convolutional network is compatible with DEAP implementations.

\begin{figure}[ht]
  \begin{center}
  \includegraphics[width=0.5\textwidth]{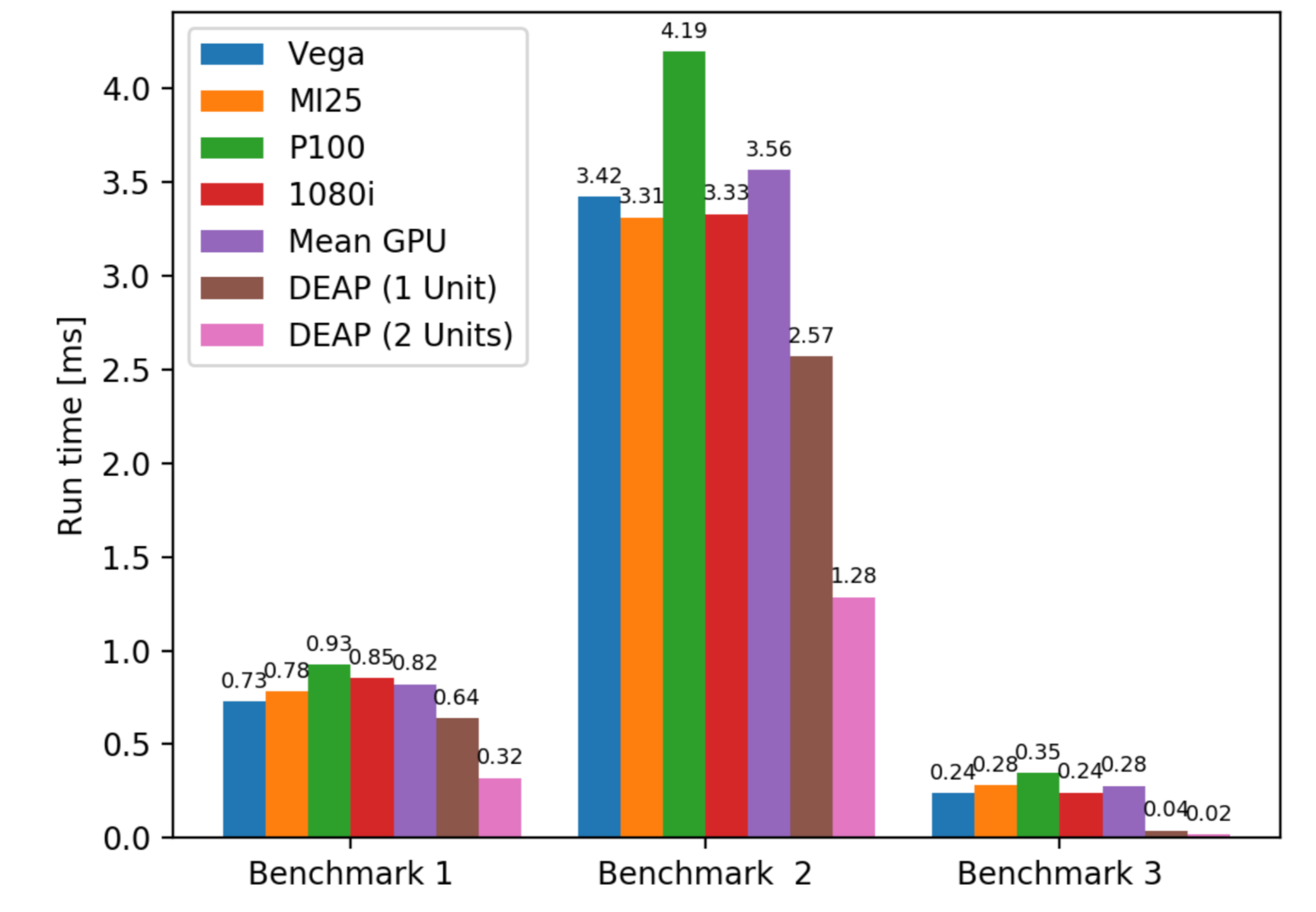}
  \caption[]{Estimated DEAP convolutional runtime compared to actual GPU runtimes from DeepBench benchmarks}
  \label{fig:deap_runtimes}
  \end{center}
\end{figure}

The estimated DEAP runtimes using one and two convolutional units were plotted against actual DeepBench runtimes in \figref{deap_runtimes}. From this, we can see that using two convolutional units performs slightly better than all the GPU benchmarks. While mean GPUs power consumption is 295 W, DEAP with a single convolutional unit sues about 110 W. Therefore, DEAP can perform convolutions between 1.4 and 7.0 faster than the mean GPU runtime while using 0.37 times the energy consumption. Using two convolutional units doubles the speed of DEAP, meaning that DEAP can be between 2.8 and 14 times faster than a conventional GPU while using almost 0.75 times the energy consumption. DEAP with a single unit performing at a speed somewhat similar to the GPUs is expected.

\section{Conclusion}
We have proposed a photonic network, DEAP, suited for convolutional neural networks. DEAP was estimated to perform convolutions between 2.8 and 14 times faster than a GPU while roughly using 0.75 times the energy consumption. A linear increase in processing speeds corresponds to a linear increase in energy consumption, allowing for DEAP to be as scalable as electronics.

High level software simulations have shown that DEAP is theoretically capable of performing a convolution. We demonstrate that our DEAP CNN is capable of solving MNIST handwritten recognition task with an overall accuracy of 98\%.
The largest bottlenecks is the I/O interfacing with digital systems via DACs and ADCs. If photonic DACs\cite{Zhang:2015} and ADCs\cite{Piqueras:2011} are to be built with higher bit-precisions, the speedup over GPUs could be even higher. If higher bit precision photonic DACs and ADCs are able to be built, replacing the electronic components with optical ones can significantly decrease the runtime.

In order to realize a physical implementation, there are a number of issues that still need to be solved. Packaging a silicon photonic with an electronic chip with high I/O count is a challenging RF engineering task, but it is a central thrust in the roadmap for silicon photonic foundries~\cite{Rahim:2018}. 
There also needs to be control circuitry that routes the outputs of the SDRAM into the relevant DACs and from the ADCs into the SDRAM. Since we assume that the control circuitry can operate significantly faster than a memory access, we believe it will have a negligible impact on the overall throughput. Another issue is that DEAP processes data in the analog domain, whereas GPUs perform floating point arithmetic. Though floating-point arithmetic does have some degree of error due to rounding in the mantissa, their errors are deterministic and predictable. On the other hand, the errors from photonics are due to stochastic shot, spectral, Johnson-Nyquist and flicker noises, as well as quantization noise in the ADC, and distortion from the RF signals applied to the modulators. However, artificially adding random noise to CNNs have been shown to reduce over-fitting \cite{You:2018}, meaning that some degree of stochastic behaviour is tolerable in the domain of machine learning problems.

Finally, MRRs have only been shown to have up to 7-bits of precision, which is significantly smaller than the range precision supported by even half-precision (16-bit) floating point representations. In conclusion, photonics has the potential to perform convolutions at speeds faster than top-of- the-line GPUs while having a lower energy consumption. Moving forward, the greatest challenges to overcome have to do with increasing the precision of photonic components so that they are comparable to classical floating-point representations. Overall, silicon photonics has the potential to outperform conventional electronic hardware for convolutions while having the ability to scale up in the future.


\section*{Acknowledgment}
Funding for B.J.S., B.A.M., H.B.M., and V.B. was provided by the Natural Sciences and Engineering Research Council of Canada (NSERC) and the Queen's Research Initiation Grant (RIG).

\bibliographystyle{apsrev4-2}
\bibliography{references.bib}

\end{document}